\documentclass[twocolumn,showpacs,preprintnumbers,amsmath,amssymb]{revtex4}
\usepackage{graphicx}% Include figure files
\input{epsf.sty}
\begin{document}
%\preprint{}

\title{Fulde-Ferrell-Larkin-Ovchinnikov state in  a perpendicular field of quasi two-dimensional CeCoIn$_5$}% Force line breaks with \\

\author{K.~Kumagai$^1$, M. Saitoh$^1$, T. Oyaizu$^1$, Y. Furukawa$^1$, S.~Takashima$^2$, M.~Nohara$^2$, H.~Takagi$^2$, and Y.~Matsuda$^{3,4}$}
\affiliation{$^1$Division of Physics, Graduate School of Science, Hokkaido University,  Sapporo 060-0810, Japan }%
\affiliation{$^2$Department of Advanced Materials Science, University of Tokyo, Kashiwanoha, Kashiwa, Chiba 277-8581, Japan}%
\affiliation{$^3$Department of Physics, Kyoto University, Kyoto 606-8502, Japan}%
\affiliation{$^4$Institute for Solid State Physics, University of Tokyo, Kashiwanoha, Kashiwa, Chiba 277-8581, Japan}%

%\date{\today}

\begin{abstract}

A Fulde-Ferrell-Larkin-Ovchinnkov (FFLO) state was previously reported in the quasi-2D heavy fermion CeCoIn$_5$ when a magnetic field was applied parallel to the $ab$-plane.   Here, we conduct $^{115}$In NMR studies of this material in a {\it perpendicular}  field, and provide strong evidence for FFLO in this case as well.    Although the topology of the phase transition lines in the $H$-$T$ phase diagram is identical for both configurations, there are several remarkable differences between them.    Compared to  $H\parallel ab$,  the FFLO region for $H\perp ab$ shows a sizable decrease, and  the critical field separating the FFLO and non-FFLO superconducting states almost ceases to have a temperature dependence.   Moreover, directing $H\perp ab$ results in a notable change in the quasiparticle excitation spectrum within the planar node associated with the FFLO transition.

\end{abstract}


\maketitle

In spin singlet superconductors, superconductivity is suppressed by a magnetic field as a consequence of its coupling to the electronic orbital angular momentum (vortices) or to the conduction electron spins (Pauli paramagnetism).   When Pauli pair-breaking dominates over the orbital effect, an exotic superconducting phase was predicted to appear by Fulde-Ferrell and Larkin-Ovchinnikov (FFLO)  at high fields and low temperatures \cite{fflo}.     In the FFLO state, pair-breaking due to the Pauli paramagnetic effect is reduced by the formation of a new pairing state ({\boldmath$k$}$\uparrow$,  {\boldmath $-k+q$}$\downarrow$) with $\mid${\boldmath $q$}$\mid$  $\sim 2 \mu_B H / \hbar v_F $ ($v_F$ is the Fermi velocity) between the Zeeman split parts of the Fermi surface, instead of ({\boldmath $k$}$\uparrow$,  {\boldmath $-k$}$\downarrow$)-pairing in ordinary BCS superconductors.   A fascinating aspect of the FFLO state is that an inhomogeneous superconducting phase with a spatially oscillating order parameter $\Delta$({\boldmath $r$})$=\Delta_0\cos (${\boldmath $q$}$\cdot${\boldmath $r$}) and spin polarization  shows up in the vicinity of the upper critical field $H_{c2}$ \cite{fflo,maki,shimahara1,yan,budgin,shimahara2,gg,adachi,maki2,tachiki,mizushima}.    It has been shown that in the presence of a substantial orbital effect,  2D planar nodes appear periodically perpendicular to the Abrikosov vortex lattice. \cite{gg,maki2,adachi,tachiki,mizushima} 

Despite the straightforward nature of the theoretical prediction, actual observations of the FFLO state have turned out to be extremely difficult.    In fact, no solid evidence,  which is universally accepted as proof of the FFLO state, has turned up in any superconductors thus far.    Recently, a new candidate for the realization of the FFLO phase has been found in heavy fermion CeCoIn$_5$ \cite{pet} with a quasi-2D electronic structure \cite{shishido}, when the magnetic field is applied {\it parallel} to the $ab$-plane ($H\parallel ab$).   Heat capacity measurements revealed a second order phase transition line within the superconducting state, indicating a new superconducting phase at the high-$H$/low-$T$ corner in the $H$-$T$ phase diagram \cite{radovan,bianch}.   At the transition line,  ultrasound velocity measurements  revealed the collapse of the flux line lattice tilt modulus \cite{watanabe}.  In addition, thermal conductivity  display a kink \cite{capan}.      Moreover, recent NMR measurements  indicate clear evidence of the spatially inhomogeneous quasiparticle structure inside the new superconducting phase \cite{kakuyanagi1}.  All of these results are precisely expected in a FFLO state.   CeCoIn$_5$ appears to meet, in an ideal way, some strict requirements placed on the existence of the FFLO state.  First, the superconducting-normal transition at $H_{c2}$ is in the first order,  indicating that superconductivity is limited by the Pauli paramagnetic effect \cite{izawa,bianch2,tayama}.    Second, it has been pointed out that the FFLO state is readily destroyed by impurities \cite{yan,adachi,takada}, but  this material is in the extremely clean regime \cite{kasahara}.  Third, the superconducting symmetry is most likely to be the $d$-wave \cite{izawa,mov,wei}, which extends the stability of the FFLO state \cite{yan,shimahara2,maki3}.
  
While there is growing experimental evidence that the FFLO state can indeed be realized in CeCoIn$_5$, several unexpected features have been found.   One of the most intriguing features is a remarkable $T$- and $H$-dependence of the phase boundary between the FFLO and non-FFLO superconducting state (hereafter referred to as the BCS state);  $H_{FFLO}^{\parallel}$ exhibits an unusually large shift to higher fields with higher temperatures \cite{radovan,bianch}.  Although a number of theories were considered to explore the FFLO state before the experimental results of CeCoIn$_5$ were reported,  none of them predicted such a large $T$-dependence of $H_{FFLO}^{\parallel}$.     Moreover, it seems to be widely believed that the 2D nature of CeCoIn$_5$ is essential for the formation of the FFLO state.  This is because both the strong reduction of the orbital pair breaking in a parallel field and the nesting properties of the quasi-2D Fermi surface are expected to stabilize the FFLO state, as discussed in layered organic superconductors \cite{shimahara1,tanatar}.   This is supported by the torque measurements that indicate the disappearance of the FFLO state when the magnetic field is slightly tilted out of the $ab$-plane \cite{radovan}.   On the other hand,  specific heat measurements show an anomaly for $H\perp ab$ just below $H_{c2}^{\perp}$ at low temperatures, similar to $H\parallel ab$ \cite{bianch}.   Thus,  whether the FFLO state can exist when $H\perp ab$ is controversial and its clarification is very important for understanding the FFLO state in CeCoIn$_5$.

We here direct our attention  to the superconducting state of CeCoIn$_5$ in a {\it perpendicular}  field ($H\perp ab$).  We present  evidence for the FFLO state  in this case as well.   We show that directing the magnetic field towards the $c$-axis results not only in a dramatic change in the FFLO region in the {\it H-T}  phase diagram, but also in a notable change in the quasiparticle excitation spectrum within the FFLO nodal sheets.

High quality single crystals of CeCoIn$_5$ with $T_c=2.3$~K were grown by a flux method.   %Specific heat measurements showed a very sharp transition at $T_c=2.3$~K, indicative of a high quality sample.    The upper critical field with  {\boldmath $H$}  parallel and perpendicular to the $ab$-plane are $H_{c2}^{\parallel}$=11.73~T and $H_{c2}^{\perp}$=4.80~T, respectively.   
$^{115}$In ($I$=9/2) NMR measurements were performed using a phase-coherent pulsed NMR spectrometer.    Experiments were always carried out in a magnetic field {\boldmath $H$}  perpendicular to the $ab$-plane under the field-cooled condition.   We reduced the rf excitation power to more than 1/100 times smaller than that used in ordinary experiments in order to avoid Joule heating.  Spectra were obtained from a convolution of Fourier  transform signals of the spin echo which were measured at  20~kHz intervals.   The tetragonal crystal structure of CeCoIn$_5$ consists of alternating layers of CeIn$_3$ and CoIn$_2$,  and this has two inequivalent  $^{115}$In sites per unit cell.    We report  NMR results at the In(1) site with axial symmetry in the CeIn$_3$ layer, which is located in the center of the square lattice of Ce atoms.   The Knight shift $^{115}K$ was obtained from the $^{115}$In-line using a gyromagnetic ratio of $^{115} \gamma$=9.3295~MHz/T,  and by taking into account the electric quadrupole interaction.%\cite{curro}. 

%\begin{figure}[b]
%\begin{center}
%	\includegraphics[clip,scale=0.5]{H4.8Tdown2.EPS}\\
%	\caption{$^{115}$In-NMR spectra of of CeCoIn$_5$ (\#2) as a function of Knight shift for various temperatures at $H$=4.8~T. The spectra shown by red and black lines are obtained in the superconducting and the normal region, respectively.}
%\end{center}
%\end{figure}

Figure 1(a) depicts the temperature evolution of the NMR spectrum that displayed as a function of the Knight shift in sample \#2 at $H$=4.80~T  just below  $H_{c2}^{\perp}$=4.95~T at $T=0$ for $H\perp ab$. %(see Fig.4~(a)).
 Upon entering the superconducting state $(T_c(H)=0.48~$K),  the NMR intensity is strongly reduced.   One immediately notices that the NMR spectrum changes dramatically below $T_c(H)$.  While the NMR spectrum is nearly symmetrical above $T_c(H)$,  the spectrum at $T=$0.46~K exhibits a shoulder structure at lower frequencies.    This shoulder structure is followed by a double peak structure, which clearly shows up below $T=$0.43~K.    The double peak structure persists down to the lowest temperature (70~mK), but the intensity of the higher resonance line decreases with decreasing temperature.   Figure 1 (b), (c) and (d) display the spectra in sample \#1, which is taken in a different batch from \#2, at 4.6~T, 4.8~T and 4.85~T, respectively.  At 4.8~T and 4.85~T,  the change in the spectrum just below $T_c(H)$ is pronounced; a distinct double peak structure is observed.   However, the double peak structure was not observed well below $T_c(H)$, though the spectrum exhibit a broadening at higher frequencies (at higher Knight shift).  On the other hand, for $H=$4.6~T,  shown in Fig. 1(b), the spectrum shifts smoothly toward the low frequency side below $T_c(H)$ without showing the double peak anomaly.

\begin{figure}[b]
\begin{center}
  \includegraphics[scale=0.40]{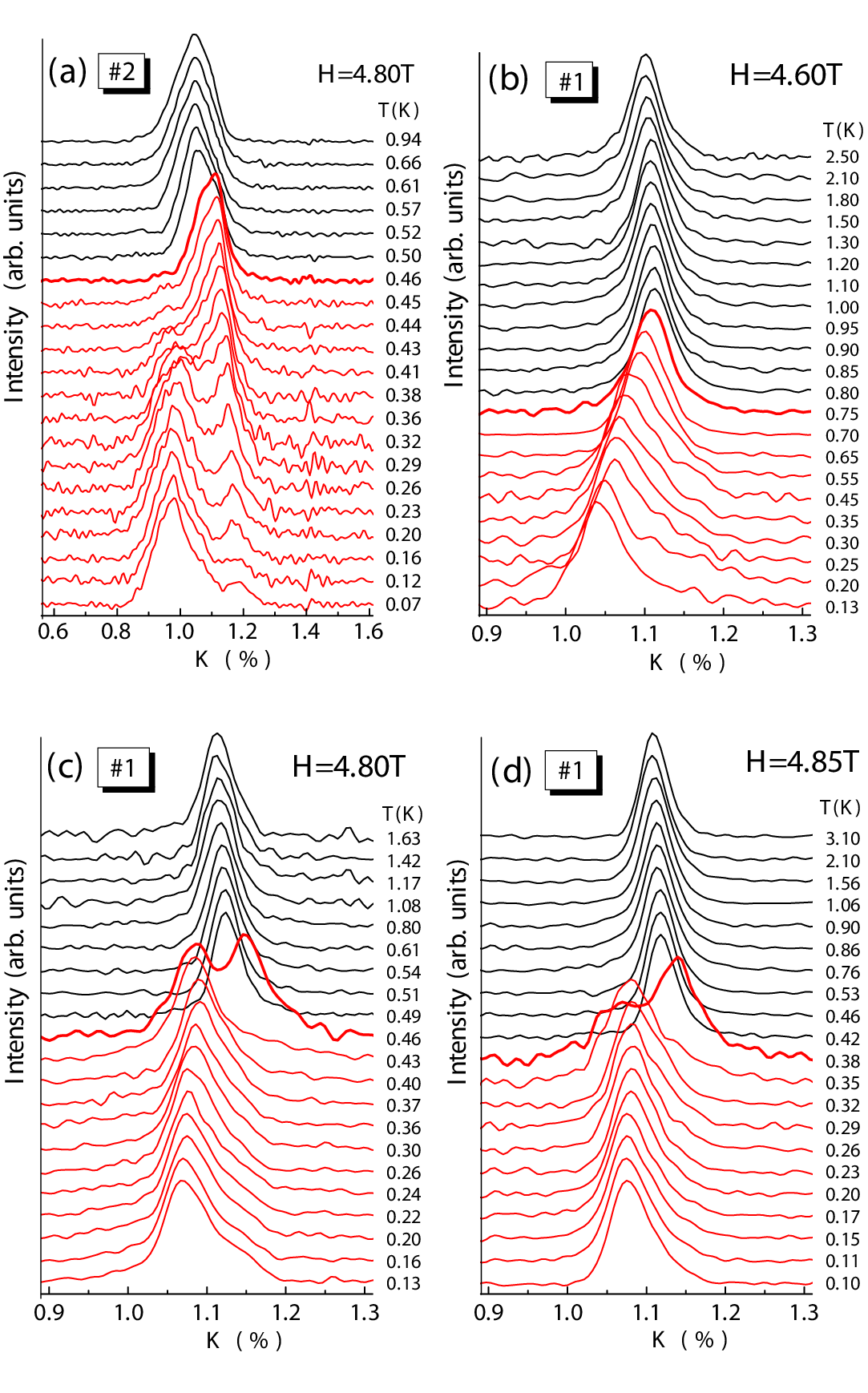}\\
\caption{$^{115}$In-NMR spectra of CeCoIn$_5$ (\#1 and \#2) as a function of the Knight shift when going from a normal state (black lines) to a superconducting state (red lines)  at (a) $H$=4.8~T (\#2), (b) 4.60~T (\#1), (c) 4.80~T (\#1) and (d) 4.85~T  (\#1).  The intensity is normalized by the largest peak intensity. }
\end{center}
\end{figure}
    
%\begin{figure}[b]
%\begin{center}
%	\includegraphics[clip,scale=0.38]{H4.8Tdown2.EPS}    
 %   \includegraphics[scale=0.38]{K_H4.60T.EPS}\\
  %  \includegraphics[scale=0.38]{K_H4.80T.EPS}
  %  \includegraphics[scale=0.38]{K_H4.85T.EPS}\\
    %\includegraphics[scale=0.36]{K_H4.90T.EPS}\\
%    \caption{$^{115}$In-NMR spectra of of CeCoIn$_5$ (\#1 and \#2) as a function of Knight shift when going from normal state (black lines) to superconducting state (red lines)  at (a): $H$=4.8~T (\#2), (b): 4.60~T (\#1), (c): 4.80~T (\#1) and (d): 4.85~T  (\#1).  }

%\end{center}
%\end{figure}

\begin{figure}[t]
\begin{center}
	\includegraphics[clip,scale=0.6]{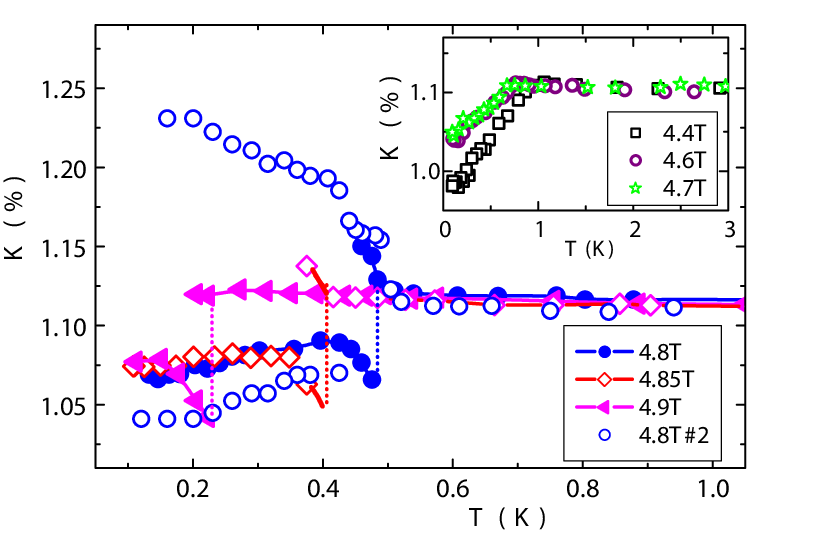}\\
	\caption{$T$-dependence of the Knight shift, $^{115}K$,  at $H$= 4.8~T (blue $\bullet$), 4.85~T (red $\diamond$), and 4.90~T (purple $\triangleleft$) for sample  \#1,  and also at $H$=4.8~T (blue $\bigcirc$) for \#2.  $T_c(H)$=0.48~K , 0.4~K, and 0.22~K at $H$=4.8~T, 4.85~T, and 4.9~T, respectively.  
The inset shows the detail of the low field region below $H=$4.7~T for  \#1. The solid and dotted lines are guide for eyes. } 
\end{center}
\end{figure}
    
The anomalous behavior of the spectrum  can also be seen clearly by plotting the Knight shift,  which is defined as the peak position of the spectrum.   Figure 2 and the inset depict the temperature dependence of the Knight shift when going across the normal-to-superconducting transition.    In the normal state, the Knight shift is nearly $T$-independent and its value is close to the one reported previously \cite{curro}.  As shown in the inset of Fig.~2,  the Knight shift at $H\leq$4.7~T decreases monotonically with decreasing temperature in the superconducting state below $T_c(H)$.  This behavior, together with the NMR spectrum shown in Fig.~1(b), are typical  in the BCS state.   As shown in the main panel of Fig.~2,  above $H$=4.7~T, the Knight shift changes dramatically below $T_c(H)$.  The Knight shift exhibits a jump at $T_c(H)$ and splits into two values.  While the Knight shift of the lower resonance line is reduced from the value in the normal state, the Knight shift of the higher resonance line is strongly enhanced.    Interestingly, for  sample \#1, the Knight shift for the lower resonance line exhibits an overshoot behavior just below $T_c(H)$.  The jump in the Knight shift, along with the overshoot behavior,  definitely indicates the first order nature of the superconducting-to-normal transition. 
 
We  measured the NMR spectrum for several crystals in  different batches, and always observed  spectra similar to the ones shown  in Fig.~1 (a) or (c)  below $T_c(H)$ in a narrow field range between $H$=4.7 to 4.9~T.  We therefore conclude that the double peak structure in the narrow region at the high-$H$/low $T$ corner in the $H$-$T$ plane is a common feature in the NMR spectrum, though the relative intensity of the peak just below $T_c(H)$ is sample-dependent.  

We stress that {\it the double peak structure and its temperature evolution in the perpendicular field bear a striking resemblance to those in the parallel field} (see Fig.~2 in Ref.\cite{kakuyanagi1}).   As seen in Fig.~2, the Knight shift of the lower resonance line is close to the Knight shift at $H$=4.6~T, indicating that the origin of the lower resonance line,  which dominates well below $T_c(H)$,  is the same as the resonance line in the BCS state.   %region with no anomalous double peak structure. 
Therefore, as discussed in Ref.\cite{kakuyanagi1},  the appearance of the higher resonance peak is a manifestation of the new  quasiparticle region that appeared as a result of spatial modulation of the superconducting order parameter in the FFLO state.  Moreover, it has been shown that the temperature evolution of the higher resonance peak can be accounted for semi-quantitatively by considering the inhomogeneous distribution of the shielding currents passing across the planar nodes that are periodically aligned perpendicular to the flux line lattice.  We note that the sample-dependent double peak structure can be explained by the distribution of the shielding currents which are influenced by the sample shape.

\begin{figure}[t]
\begin{center}
	\includegraphics[scale=0.60]{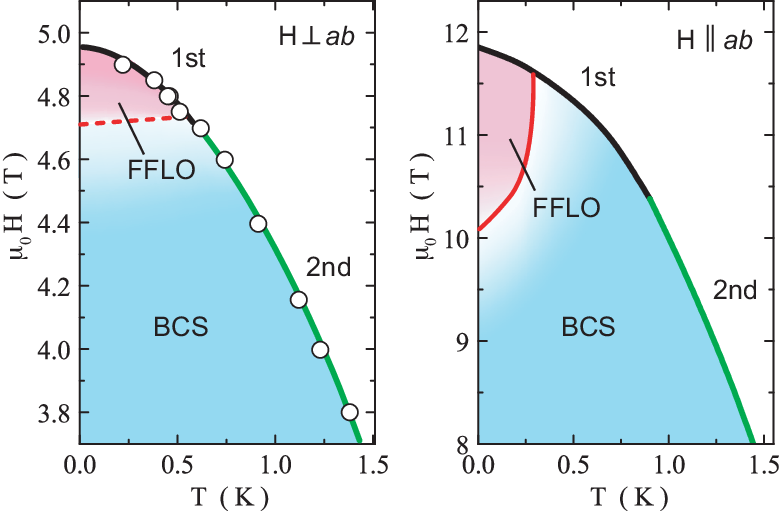}\\
	\caption{$H$-$T$ phase diagrams at low temperatures and high fields  for {\boldmath $H$}$\perp ab$ (left) and for $H\parallel ab$ (right).  The colored portions display the FFLO (pink) and BCS (blue) regions.    The open circles are $H_{c2}^{\perp}$ determined by the NMR experiments.  The black and green lines represent the upper critical fields which are in the first order and in the second order, respectively.    The red dashed and solid lines represent the phase boundary separating the FFLO and BCS states.@For  {\boldmath $H$}$\perp ab$,  precise determination of the phase boundary between the FFLO and the BCS states is difficult by the present experiments. }
\end{center}
\end{figure}	

Thus,  the present NMR data in this study lead us to conclude that the FFLO state is indeed realized for $H\perp ab$.  The occurrence of the first order transition at  $H_{c2}^{\perp}$ implies that the Pauli paramagnetic effect dominates near $H_{c2}^{\perp}$ in this case as well.   We first argue  the requirement for the FFLO formation with respect to the Pauli effect for $H\perp ab$.  Such a requirement can be quantitatively described by the Maki parameter $\alpha=\sqrt{2}H_{orb}/H_P$  \cite{maki}, which is the ratio of the orbital limiting upper critical field $H_{orb}$ and the Pauli limiting field $H_{P}$.  It has been shown that the required minimum value of the Maki parameter for the formation of the FFLO state is $\alpha$=1.8 \cite{gg}.   The orbital limited fields obtained from the initial slope of $H_{c2}$, $dH_{c2}/dT  |_{T=T_c}$ through the relation $H_{orb}=-0.7T_c dH_{c2}/dT$,  are 38.6~T and 17.7~T for $H \parallel ab$ and $H \perp ab$, respectively.   Assuming $H_P\simeq H_{c2}(0)$, the Maki parameters for $H\parallel ab$ and $H\perp ab$ are estimated to be $\alpha^{\parallel}$=4.6 and $\alpha^{\perp}$=5.0, respectively.  Thus $\alpha^{\perp}$ greatly exceeds the minimum value of 1.8 for the FFLO state.   We note that $\alpha^{\perp}$ is even larger than $\alpha^{\parallel}$.    The  extremely large $\alpha^{\perp}$ can be attributed to the strongly enhanced Pauli paramagnetic susceptibility in the normal state.   In fact,  the Pauli susceptibility for $H\perp ab$ is nearly twice as large as that for $H\parallel ab$ ~\cite{bianch}.

We next discuss the FFLO region in the $H$-$T$ phase diagram.  The absence of the double peak anomaly in the spectrum below  $H=4.7$~T indicates that the critical field $H_{FFLO}^{\perp}$ separating the FFLO and the BCS state lies in a very narrow range between 4.7~T and 4.75~T below $T_c(H)$,  though its precise $T$-dependence is difficult to determine.   Figure~3 (a)  displays the $H$-$T$ phase diagram for $H\perp ab$ as determined by the present measurements, while Fig.~3(b)  displays the diagram for $H\parallel ab$ reported previously.~\cite{bianch,capan}  It should be noted that the specific heat measurements showed an anomaly in the vicinity of  $H_{FFLO}^{\perp}$.~\cite{bianch}  Obviously, the topology of the phase transition lines for both configurations is identical in that the FFLO line branches from the first order superconducting-to-normal transition line.   

Despite the striking resemblance in the phase diagrams of both configurations,  there are distinguishing differences between them.   For  $H\perp ab$, the FFLO phase occupies a tiny high-$H$/low-$T$ corner in the $H$-$T$ phase diagram, compared to the case for $H\parallel ab$.   In addition,  $H_{FFLO}^{\perp}$ almost ceases to have a temperature dependence, in striking contrast to $H\parallel ab$.   Moreover, for $H\perp ab$,  the end point of the FFLO transition line appears  close to the end point of the first order transition, though the end point of the first order transition line is difficult to determine explicitly by the experiments.  Most theories that attempt to describe the FFLO state have predicted that the critical field separating the FFLO and BCS states is nearly temperature independent.  In this respect, the phase diagram for $H\perp ab$ seems to be close to the phase diagram of  {\it conventional}  FFLO theories, although such a simple comparison should be scrutinized.   

In addition to the shrinkage of the FFLO region, the directing $H$ towards the $c$-axis also results in a notable change in the quasiparticle excitation spectrum.   We can examine the quasiparticle structure inferred from the NMR spectrum.   The well separated double resonance lines provide two important pieces of  information for the quasiparticle excitations in the FFLO state.   First, {\it the quasiparticles excited around the FFLO planar nodes are spatially well-separated from those excited around vortex cores.}   According to  recent theory,  the quasiparticles are not excited in the region where the vortex lines intersect  the planar node,  because of the bound states due to the $\pi$ shift of the pair potential associated with the planar node \cite{mizushima}.  This indicates that the quasiparticle regions within the vortex line do not spatially overlap with those in the FFLO planar nodes.   Second,  the Knight shift of the higher resonance line is strongly enhanced from the normal state value, as shown by open circles in the main panel of Fig.~2.   This immediately indicates that {\it the quasiparticle state within the planar nodes created in the FFLO state for $H\perp ab$ should be distinguished from the normal state}.  The strongly enhanced Knight shift for $H\perp ab$  indicates that the paramagnetic moments are enhanced within the nodal sheets.  The results for $H\perp ab$ are in sharp contrast with the results for $H\parallel ab$, wherein the Knight shift of the higher resonance line coincides with the normal state value at very low temperature. Thus, the quasiparticles in the planar nodes in the case of  $H\parallel ab$ are in a state similar to the normal state \cite{kakuyanagi1}.  

We finally discuss the difference between $H\parallel ab$ and $H \perp ab$.     The sizable decrease of the FFLO region for $H\perp ab$ indicates that the FFLO state is more stable for $H\parallel ab$.   This may be related to the shape of the Fermi surface with a quasi-2D nature, though the origin of the remarkable $T$-dependence of $H_{FFLO}^{\parallel}$ is unknown.    At present, the origin of the enhancement of the paramagnetic moment in the FFLO sheets observed only for  $H\perp ab$ is also an open  question.   We stress that the role of antiferromagnetic fluctuations near the quantum critical point, which is important for the superconductivity and the non-Fermi liquid behavior in the normal state of CeCoIn$_5$ \cite{qcp},  on the FFLO state is not well understood so far.   In fact, very recent heat capacity measurements under pressure clearly demonstrate that antiferromagnetic fluctuations is unfavorable for the stability of the FFLO state \cite{miclea}.   How the antiferromagnetic fluctuations affect  both the FFLO phase diagram and the quasiparticle excitation spectrum in the FFLO state calls for further investigations.

To summarize,  $^{115}$In NMR studies in CeCoIn$_5$ revealed the presence of the FFLO state in a {\it perpendicular} magnetic field  as well.   The FFLO phase diagram in a perpendicular field bears resemblance to that in a parallel field.   However,  several notable differences,  not only in the FFLO  phase diagram but also in the quasiparticle excitations within the FFLO planar nodes, were found between the two different configurations. 

We thank M.~Ichioka, R.~Ikeda, K.~Machida, K.~Maki, and H.~Shimahara for valuable discussions.

\end{document}